\documentclass{ws-procs9x6}

\usepackage{units}
\usepackage{upgreek}

\newcommand{\refeq}[1]{(\ref{#1})}
\def\etal {{\it et al.}}

\begin{document}

\title{OBTAINING BOUNDS FROM\\
ULTRA-HIGH ENERGY COSMIC RAYS IN\\
ISOTROPIC MODIFIED MAXWELL THEORY}

\author{M.\ SCHRECK}

\address{ANKA Storage Ring, Karlsruhe Institute of Technology \\
76344 Eggenstein-Leopoldshafen, Germany \\
E-mail: Marco.Schreck@kit.edu}

\begin{abstract}
This article reviews the methods used to obtain a two-sided bound on isotropic modified Maxwell theory from experimental data of ultra 
high-energy cosmic rays in 2008. The bound is updated with results from the HEGRA experiment.
\end{abstract}

\bodymatter

\section{Introduction}

The advent of quantum mechanics at the beginning of the 20th century marks an important step into a new era of physics. The classical
theory was left behind and it became possible to describe physical processes occurring at the atomic length scale of $\unit[10^{-10}]{m}$.
Quantum mechanics was then also applied to atomic nuclei reducing the length scale of its applicability again by five orders of magnitude.
In the aftermath, quantum field theory was developed, which for several decades has been used to great success in order to understand
physics down to approximately $\unit[10^{-18}]{m}$.

Perhaps a similar revolution is currently taking place at the beginning of the 21st century. The interest in experiments looking for signs of
quantum gravity, which is expected to play a role at the Planck length i.e. at $\unit[10^{-35}]{m}$, has been steadily increasing. If
Einstein's relativity and quantum physics are assumed to be still valid at this scale, both theories will have to merge into a new theory 
correctly describing fluctuations of spacetime itself. Currently there is no chance of investigating quantum gravitational effects directly 
at the Planck energy. However there is a clear signal for such effects that may be visible far below the Planck energy: a violation of 
Lorentz symmetry.

The latter is motivated by various approaches to a fundamental theory: string theory, loop quantum gravity, noncommutative spacetime, etc.
Since we currently still do not know how such a theory looks like it makes sense to study Lorentz violation within a model-independent
framework: the Standard-Model Extension (SME) \cite{ColladayKostelecky1998}. This is a collection of all field theoretic operators
that are symmetric with respect to the gauge group $\mathit{SU}_{\mathrm{c}}(3) \times \mathit{SU}_L(2) \times \mathit{U}_Y(1)$ of the ordinary
Standard Model but that violate particle Lorentz invariance. The current goal of experiments is to measure the parameters associated with these
operators. A detection of nonvanishing Lorentz-violating coefficients could deliver significant insights on how different particle sectors are
indeed affected by quantum gravitational phenomena. So far it has been impossible to detect Lorentz violation. Hence with better and better
experiments even stricter constraints on Lorentz-violating parameters can be set.

\section{Modified Maxwell theory}

In the power-counting renormalizable photon sector of the SME there exist two modifications: Maxwell--Chern--Simons (MCS) theory and modified
Maxwell theory. The first is characterized by a dimensionful scale and a background vector field where the second involves a dimensionless
fourth-rank tensor background field. The dimensionful parameter of MCS theory was already heavily bounded by astrophysical observations
\cite{Carroll-etal1990} whereas some of the parameters of modified Maxwell theory --- especially the isotropic sector --- were only weakly 
bounded in 2008. This was the motivation to consider a modified quantum electrodynamics (QED) in Ref.\ \refcite{Klinkhamer:2008ky} that 
results from minimally coupling modified Maxwell theory to a standard Dirac theory of spin-1/2 fermions with charge $e$ and mass $M$:
\begin{subequations}
\begin{equation}
S_{\mathrm{modQED}}=S_{\mathrm{modMax}}+S_{\mathrm{standDirac}}\,,
\end{equation}
\begin{equation}
\label{eq:modified-maxwell-theory}
S_{\mathrm{modMax}}=\int_{\mathbb{R}^4} \mathrm{d}^4x\,\left\{-\frac{1}{4}F^{\mu\nu}(x)F_{\mu\nu}(x)-\frac{1}{4}\kappa^{\mu\nu\varrho\sigma}F_{\mu\nu}(x)F_{\varrho\sigma}(x)\right\}\,,
\end{equation}
\begin{equation}
\label{eq:standard-dirac-theory}
S_{\mathrm{standDirac}}=\int_{\mathbb{R}^4} \mathrm{d}^4x\,\overline{\psi}(x)\Big\{\gamma^{\mu}\big[\mathrm{i}\partial_{\mu}-eA_{\mu}(x)\big]-M\Big\}\psi(x)\,.
\end{equation}
\end{subequations}
Here $F_{\mu\nu}(x)\equiv \partial_{\mu}A_{\nu}(x)-\partial_{\nu}A_{\mu}(x)$ is the field strength tensor of the \textit{U}(1) photon field $A_{\mu}(x)$ 
and $\psi(x)$ is the spinor field. The fields are defined in Minkowski spacetime with the metric tensor $(\eta_{\mu\nu})=\mathrm{diag}(1,-1,-1,-1)$.
The second term of Eq.\ \refeq{eq:modified-maxwell-theory} containing the fourth-rank tensor background field $\kappa^{\mu\nu\varrho\sigma}$
manifestly violates particle Lorentz invariance. This field is a low energy effective description of possible physics at the Planck scale
and it defines preferred spacetime directions.

The modified photon action given by Eq.\ \refeq{eq:modified-maxwell-theory} can be restricted to a sector exhibiting nonbirefringent\footnote{at
first order in Lorentz violation} photon dispersion laws. This is possible via the following ansatz for the background field:
\begin{equation}
\kappa^{\mu\nu\varrho\sigma}=\frac{1}{2}(\eta^{\mu\varrho}\widetilde{\kappa}^{\nu\sigma}-\eta^{\mu\sigma}\widetilde{\kappa}^{\nu\varrho}
-\eta^{\nu\varrho}\widetilde{\kappa}^{\mu\sigma}+\eta^{\nu\sigma}\widetilde{\kappa}^{\mu\varrho})\,,
\end{equation}
where $\widetilde{\kappa}^{\mu\nu}$ is a symmetric and traceless $(4\times 4)$-matrix: $\widetilde{\kappa}^{\mu\nu}=\widetilde{\kappa}^{\nu\mu}$, $\widetilde{\kappa}^{\mu}_{\phantom{\mu}\mu}=0$. The isotropic part of modified Maxwell theory is then defined by the choice
\begin{equation}
(\widetilde{\kappa}^{\mu\nu})\equiv \frac{3}{2}\widetilde{\kappa}_{\mathrm{tr}}\,\mathrm{diag}\left(1,\frac{1}{3},\frac{1}{3},\frac{1}{3}\right)\,.
\end{equation}
From the field equations results a modified photon dispersion law that is isotropic in three-space:
\begin{equation}
\label{eq:modified-dispersion-law}
\omega(k)=c\,\sqrt{\frac{1-\widetilde{\kappa}_{\mathrm{tr}}}{1+\widetilde{\kappa}_{\mathrm{tr}}}}k\,.
\end{equation}
Here $\omega$ is the photon frequency and $k$ the wave number. Note that $c$ is the maximum attainable velocity of the Dirac particle
described by Eq.\ \refeq{eq:standard-dirac-theory}, which is not affected by Lorentz violation. Dependent on the choice of
$\widetilde{\kappa}_{\mathrm{tr}}$ the modified photon velocity does not coincide with the maximum velocity of massive particles any
more. This leads to peculiar particle physics processes, which are forbidden in standard QED.

\section{Vacuum Cherenkov radiation and photon decay}

The modified photon dispersion relation of Eq.\ \refeq{eq:modified-dispersion-law} allows $\widetilde{\kappa}_{\mathrm{tr}}$ to lie in the interval
$(-1,1]$. For $\widetilde{\kappa}_{\mathrm{tr}}\in (0,1]$ the photon velocity is smaller than $c$ and Dirac particles can travel faster
than photons. If the latter is the case, a Cherenkov-type process in vacuum takes place leading to energy loss of the Dirac particle by the emission
of modified photons $\widetilde{\upgamma}$. This process occurs above a certain threshold energy $E_{\mathrm{thresh}}$ of the massive particle
(e.g. a proton $\mathrm{p^+}$) and its radiated energy rate $\mathrm{d}W/\mathrm{d}t$ far above the threshold is:
\begin{equation}
\label{eq:threshold-cherenkov}
E_{\mathrm{thresh}}^{\mathrm{p^+}\rightarrow \mathrm{p^+}\widetilde{\upgamma}}=\frac{M_{\mathrm{p}}c^2}{2}\sqrt{2+\frac{2}{\widetilde{\kappa}_{\mathrm{tr}}}}\,,\quad
\left.\frac{\mathrm{d}W}{\mathrm{d}t}\right|^{\mathrm{p^+}\rightarrow \mathrm{p^+}\widetilde{\upgamma}}_{E\gg E_{\mathrm{thresh}}^{\mathrm{p^+}\rightarrow \mathrm{p^+}\widetilde{\upgamma}}}\simeq \frac{7}{12}\alpha\widetilde{\kappa}_{\mathrm{tr}}\frac{E^2}{\hbar}\,.
\end{equation}
Herein, $M_{\mathrm{p}}$ is the proton mass, $\alpha\equiv e^2/(4\pi\varepsilon_0\hbar c)$ the electromagnetic fine structure constant,\footnote{with the
vacuum permittivity $\varepsilon_0$} and $\hbar$ Planck's constant. For vanishing isotropic parameter $\widetilde{\kappa}_{\mathrm{tr}}$ the threshold
energy goes to infinity and the radiated energy rate vanishes showing that the process is forbidden in standard~QED.

For $\widetilde{\kappa}_{\mathrm{tr}}\in (-1,0)$ the modified photon velocity is larger than the maximum velocity of Dirac
particles. In this case a photon may decay preferably into an electron positron pair $\mathrm{e^+e^-}$. This
decay is possible above a certain threshold energy $E_{\mathrm{thresh}}$ for the photon and it has the following decay rate~$\Gamma$:
\begin{equation}
\label{eq:threshold-photon-decay}
E_{\mathrm{thresh}}^{\widetilde{\upgamma}\rightarrow \mathrm{e^+e^-}}=M_{\mathrm{e}}c^2\sqrt{2-\frac{2}{\widetilde{\kappa}_{\mathrm{tr}}}}\,,\quad \Gamma^{\widetilde{\upgamma}\rightarrow \mathrm{e^+e^-}}|_{E\gg E_{\mathrm{thresh}}^{\widetilde{\upgamma}\rightarrow \mathrm{e^+e^-}}}\simeq -\frac{2}{3}\alpha\widetilde{\kappa}_{\mathrm{tr}}E\,,
\end{equation}
with the electron mass $M_{\mathrm{e}}$.
The crucial difference from vacuum Cherenkov radiation is the minus sign appearing together with $\widetilde{\kappa}_{\mathrm{tr}}$. This tells us that
the above equations only make sense for negative Lorentz-violating parameters.

\begin{table}[t!]
\tbl{Ultra-high energy photon and cosmic ray event. For details on the Auger event consult Ref.\ \refcite{Klinkhamer:2008ky} and references
therein. The HEGRA event is obtained from the last bin of Fig.\ 3 in Ref.\ \refcite{Aharonian:2004gb}, which has a significance of 2.7$\sigma$.
Since the energy is high enough and the source was identified as the Crab nebula, the left-hand bin endpoint with
$E\simeq\unit[56]{TeV}$ can be taken as a fiducial photon event. For the energy uncertainty a conservative estimate of 10\% is used far above
the detector threshold where the uncertainty mainly originates from statistical~errors.
}
{\begin{tabular}{@{}cccc@{}}
\toprule
Experiment   & Observation & Energy $E$          & Energy uncertainty $\Delta E/E$ \\
\colrule
HEGRA        & 1997--2002  & $\unit[56]{TeV}$ [Fig.\ 3 of Ref.\ \refcite{Aharonian:2004gb}] & 10\% [p.\ 12 of Ref.\ \refcite{Aharonian:2004gb}] \\
Auger        & ID 737165   & $\unit[212]{EeV}$ [see Ref.\ \refcite{Klinkhamer:2008ky}]      & 25\% [see Ref.\ \refcite{Klinkhamer:2008ky}]      \\
\botrule
\end{tabular}}
\label{tab:uhecr-events}
\end{table}%

\section{Updated two-sided bound on the isotropic parameter}

If a hadronic primary or a photon is detected on Earth its energy must be smaller than the threshold energy of
Eq.\ \refeq{eq:threshold-cherenkov} and Eq.\ \refeq{eq:threshold-photon-decay}, respectively. Using then the events of
Table~\ref{tab:uhecr-events} we obtain the following updated two-sided bound on the isotropic parameter of modified Maxwell theory at
the 2$\sigma$~level:
\begin{equation}
-2 \cdot 10^{-16} < \widetilde{\kappa}_{\mathrm{tr}} < 6\cdot 10^{-20}\,.
\end{equation}
The lower bound has been improved by a factor of 4 in comparison to Ref.~\refcite{Klinkhamer:2008ky}.

\section*{Acknowledgments}

It is a pleasure to thank R. Wagner for helpful discussions and for pointing out Ref.\ \refcite{Aharonian:2004gb} during the CPT'13
conference in Bloomington, Indiana.

\end{document}